\documentclass[traditabstract]{aa}
\usepackage{graphicx}
\usepackage{rotating,subfigure,amssymb,afterpage}
\usepackage{txfonts}
\usepackage{natbib}
\usepackage[pdftitle={Local Underdensity in the{\sf REFLEX II} Survey},
pdfauthor={H. B\"ohringer et al.},
pdfsubject={}, pdfkeywords={}, bookmarks, bookmarksopen, 
colorlinks=true, linkcolor=blue, citecolor=blue, anchorcolor=blue,
urlcolor=blue]{hyperref} 
\newcommand{\propsim}{\lower 3pt \hbox{$\, \buildrel {\textstyle
      \propto}\over {\textstyle \sim}\,$}}
\begin{document}

   \title{Observational evidence for a local underdensity in the Universe
and its effect on the measurement of the Hubble Constant
\thanks{
   Based on observations at the European Southern Observatory La Silla,
   Chile and the German-Spanish Observatory at Calar Alto}}

   \author{Hans B\"ohringer\inst{1}, Gayoung Chon\inst{1}, Chris A. Collins\inst{2}}

   \offprints{H. B\"ohringer, hxb@mpe.mpg.de}

   \institute{$^1$ University Observatory, Ludwig-Maximilians-Universit\"at M\"unchen,
                  Scheinerstr. 1, 81679 M\"unchen, Germany.
\\
              $^2$ Astrophysics Research Institute, Liverpool John Moores University, 
                    IC2, Liverpool Science Park, 146 Brownlow Hill,
                    Liverpool L3 5RF, UK
}

   \date{Submitted 29/7/19}

\abstract{For precision cosmological studies it is important to know the 
local properties of the reference point from which we observe the Universe.
Particularly for the determination of the Hubble constant with
low-redshift distance indicators, the values observed depend on the 
average matter density within the distance range covered. 
In this study 
we used the spatial distribution of galaxy clusters to map the 
matter density distribution in the local Universe. The study is based
on our {\sf CLASSIX} galaxy cluster survey, which is highly complete and well 
characterised, where galaxy clusters are detected by their X-ray emission.
In total, 1653 galaxy clusters outside the `zone of avoidance' fulfil
the selection criteria and are involved in this study. 
We find a local underdensity in the cluster distribution of about 
30 - 60\% which extends about 85 Mpc to the north and $\sim 170$ Mpc to the 
south. We study the density distribution as a function of redshift in detail
in several regions in the sky. For three regions for which the galaxy 
density distribution has previously been studied, 
we find good agreement between the density
distribution of clusters and galaxies. Correcting for the bias in the 
cluster distribution
we infer an underdensity in the matter distribution of about $-30 \pm 15\%$
($-20 \pm 10\%$) in a region with a radius of about 100 
($\sim 140$) Mpc. Calculating the probability of
finding such an underdensity through structure formation theory in a 
$\Lambda$CDM universe with concordance cosmological parameters, we find
a probability characterised by $\sigma$-values of $1.3 - 3.7$. This indicates 
low probabilities, but with values of around 10\% at the lower uncertainty limit,
the existence of an underdensity cannot be ruled out.
Inside this underdensity, the observed Hubble parameter will be larger by
about $5.5^{+2.1}_{-2.8} \%$, 
which explains part of the discrepancy between
the locally measured value of $H_0$ compared to the value of the Hubble
parameter inferred from the Planck observations of cosmic microwave
background anisotropies. If distance indicators outside
the local underdensity are included, as in many modern analyses, this effect is diluted. 
}


\authorrunning{B\"ohringer et al.}
\titlerunning{Underdensity in the local Universe}
   \maketitle
%

\section{Introduction}

As an integral part of the cosmic large-scale structure, galaxy clusters
are reliable tracers of the underlying dark matter distribution. Since they 
form the largest peaks in the initially random Gaussian density fluctuation 
field, and their density distribution can be statistically closely related 
to the matter density distribution
(e.g. Bardeen et al. 1986). Cosmic
structure formation theory has shown that the ratio of the cluster density
fluctuation amplitude is biased with respect to the matter density fluctuations
in the sense that the cluster density fluctuations have a larger variance.
The ratio of the RMS amplitude of the cluster density to that of the dark
matter, which is referred to as bias, is practically independent of scale 
(e.g. Kaiser 1986, Mo \& White 1996, Sheth \& Tormen 1999, Tinker et al. 2010). 

We have already found good observational support for this concept with our galaxy 
cluster surveys (B\"ohringer et al. 2000, 2004, 2013, 2017a).
We showed that the density fluctuation power spectrum of galaxy
clusters is an amplified version of the power spectrum of galaxies and of
the inferred power spectrum of the underlying dark matter distribution, where
the bias is dependent on the lower cluster mass limit exactly as predicted from 
theory (Balaguera-Antolinez et al. 2010, 2011). We further demonstrated
with simulations that the cluster density in local overdensities follows
the matter distribution. This was shown with superstes-clusters, superclusters
that were constructed such that they would collapse in the future (Chon
et al. 2015).

In this paper we exploit this property of galaxy clusters to study the 
matter density distribution in the local Universe. 
For the study we used our {\sf CLASSIX} 
(Cosmic Large-Scale Structure in X-rays) galaxy
cluster survey, the combination of the {\sf REFLEX} and {\sf NORAS} surveys
(B\"ohringer et al. 2000, 2004, 2013, 2017a), plus an extension into 
the `zone of avoidance'. This data set constitutes the most complete
and well characterised galaxy cluster sample in the nearby Universe
allowing a sufficiently dense sampling of the clusters to map the
cluster density distribution. 

There has been increasing interest in understanding the density
distribution in the local Universe, because the properties 
of the local reference point from which we
observe the Universe are important for conducting 
cosmological precision measurements. This is most apparent for measurements of the
Hubble constant performed with local distance standards. Historically,
when evidence for an accelerating universe came from observations of
distant type Ia supernovae  (SNIa; Perlmutter et al. 1999, Schmidt et al. 1998), 
models with local voids were considered as an alternative explanation 
of the SN data without dark energy or a cosmological
constant (e.g. C\'el\'erier 2000, Tomita, 2000, 2001, 
Alexander et al. 2009, February et al. 2010
and references therein). A minimum void model would require a
void size of at least about $200 h_{100}^{-1}$ Mpc with a mean
mass density deficiency of $\sim 40\%$ to explain the SN data 
in a universe without a cosmological constant (e.g. Alexander et al. 2009).
Today, with more precise SN data filling the redshift range very densely,
this void model mimicking an accelerated universe can be ruled out
(e.g. Kenworthy et al. 2019). Such void models have also been critically
discussed by Moss et al. (2011) and Marra et al. (2013).
In addition, our previous study on the cluster distribution in 
the {\sf REFLEX II} survey ruled out such a large spherical local void
(B\"ohringer et al. 2015).

The debate over the discrepancy between the Hubble constant measured locally
of about 74.0 ($\pm 1.4$) km s$^{-1}$ Mpc$^{-1}$ (e.g. Riess et al. 2018a,b, 2019)
and the value inferred from the {\sf Planck} survey of 67.4 ($\pm 0.5$) km s$^{-1}$ 
Mpc$^{-1}$ (Planck Collaboration 2016, 2018) has kept alive the discussion
about a local underdensity (e.g. Riess et al. 2018b, 2019, 
Shanks et al. 2018, 2019a,b). 
If our local cosmic neighbourhood has less
than the mean cosmic density, then the Hubble constant observed locally
is larger than that measured on large scale. 

Different tracers have been used to study the local density distribution.
Using SNIa, Zehavi et al. (1998) and Jha et al. (2007) claimed the 
detection of a local underdensity, while  
Hudson et al. (2004) and Conley et al. (2007) do not find such evidence.
Giovanelli et al. (1999) characterised the local Hubble flow out to 
200 $h^{-1}$ Mpc with galaxy clusters and find hardly any variations.
Huang et al. (1997), Frith et al. (2003, 2006), 
Busswell et al. (2004), and Keenan et al. (2013) found a local
underdensity in the galaxy distribution. In a more recent study
Whitbourn \& Shanks (2014) traced the galaxy
distribution in three larger regions, in the South Galactic Cap (SGC), 
the southern part of the North Galactic Cap (NGC), and
the northern part of the NGC using 2MASS K-band magnitudes in
connection with 6dFRGS, GAMA, and SDSS spectroscopic data out
to $z = 0.1$. They 
find a large underdense region with a deficit of about 40\% 
inside a radius of $150 h^{-1}$ Mpc in the SGC, no
deficit in the southern part of the NGC, and a less 
pronounced underdensity in the NGC north of the
equator. 

While most of these studies cover only a limited region of the sky,
{\sf CLASSIX} allows us to study the local density distribution 
over most of the sky area. In our previous study based on the 
{\sf REFLEX II} survey we found evidence for a southern
underdensity out to about 170 Mpc (B\"ohringer et al. 2015).
Here we studied the entire
extragalactic sky to investigate the local density distribution.
We also explore the diagnostics and systematic errors in more detail.  
 
The paper is organised as follows. In section 2 we give a brief description
of the survey and its characteristics and explain our method in section 3. 
In section 4.1 we explore the local underdensity monopole, show
results for different hemispheres in section 4.2, and for particular regions 
in section 4.4.
In section 4.3 we study cumulative density distributions of clusters
and derive the distribution of matter. We discuss the results
in section 5, and section 6 provides a summary and conclusion. 
Several technical points are explained in the Appendix. 
For the determination of all parameters that depend on distance we use
a flat $\Lambda$CDM cosmology with the parameters $H_0 = 70$ km s$^{-1}$
Mpc$^{-1}$ and $\Omega_m = 0.3$. Exceptions are results quoted
from the literature, for which the scaling is given explicitly.

\section{The CLASSIX galaxy cluster survey}

This study requires a cluster sample that traces the local Universe  
sufficiently densely, is statistically highly complete, and has a 
well-known selection function.
The best data base is at this moment our {\sf CLASSIX} galaxy cluster catalogue
(B\"ohringer et al. (2016).
It is the combination of our surveys in the southern sky,  {\sf REFLEX II}
(B\"ohringer et al. (2013), and the northern hemisphere, {\sf NORAS II}
(B\"ohringer et al. (2017a). Together they cover 8.26 ster of the sky
at Galactic latitudes $|b_{II}| \ge 20^o$
and the cluster catalogue contains 1773 members (of which 1653 are used here).
In this study we did not excise the regions of the Magellanic Clouds
or the VIRGO cluster (except when explicitly noted). In the completed
survey we find no
significant deficit in the cluster density in these sky areas.
We also use an extension of CLASSIX to lower Galactic latitudes into the
zone of avoidance. This region is restricted to the area with an
interstellar hydrogen column density $n_H \le 2.5 \times 10^{21}$ cm$^{-2}$,
because in regions with higher column density, X-rays are strongly absorbed
and the sky usually has a high stellar density, making the detection
of clusters in the optical extremely difficult. The values for the
interstellar hydrogen column density are taken from the 21cm survey of Dickey
\& Lockman (1990)~\footnote{We compared the
interstellar hydrogen column density compilation by Dickey \& Lockman (1990)
with the more recent data set of the 
Bonn-Leiden-Argentine 21cm survey (Kalberla et al. 2005)
and found that the differences relevant for us are of the order of at most 1\%.
Because our survey has been constructed with a flux cut based
on the Dickey \& Lockman results, we keep the older hydrogen column density
values for consistency.}.
This area amounts to another
2.56 ster and altogether the survey data cover 86.2\% of the sky. The spectroscopic
follow-up to obtain redshifts for this part of the survey is only about 70\%
complete and furthermore the completeness of the cluster sample is not as high as for 
{\sf REFLEX} and {\sf NORAS}. The cluster density we show for the 
zone of avoidance is therefore a lower limit.

The {\sf CLASSIX} galaxy cluster survey and its extension is based on 
the X-ray detection of galaxy clusters in the ROSAT All-Sky Survey
(RASS, Tr\"umper 1993, 
Voges et al. 1999). The source detection for the survey, 
the construction of the survey, and 
the survey selection function  as well as tests of the completeness of the
survey are described in B\"ohringer et al. (2013, 2017a). In summary, the 
nominal unabsorbed flux limit for the galaxy cluster detection in the RASS is
$1.8 \times 10^{-12}$ erg s$^{-1}$ cm$^{-2}$ in the
0.1 - 2.4 keV energy band. For the assessment of the large-scale structure
in this paper we apply an additional cut
on the minimum number of detected source photons of 20 counts. This has
the effect that the nominal flux limit quoted above is only reached in about
80\% of the survey. In regions with lower exposure and higher interstellar
absorption, the flux limit is accordingly higher 
(see Fig.\ 11 in B\"ohringer et al. 2013 and Fig.\ 5 in B\"ohringer et al. 2017a). 
This effect is modelled and
taken into account in the survey selection function.

We have already demonstrated with the {\sf REFLEX I} survey 
(B\"ohringer et al. 2004) that clusters provide a precise means to 
obtain a census of the cosmic large-scale matter distribution
through for example the correlation function (Collins et al. 2000), 
the power spectrum (Schuecker et al. 2001, 2002, 2003a, 2003b), 
Minkowski functionals, (Kerscher et al. 2001),
and, using {\sf REFLEX II}, with the study
of superclusters (Chon et al. 2013, 2014) and the cluster power 
spectrum (Balaguera-Antolinez et al. 2011, 2012). 
The fact that clusters follow the large-scale
matter distribution in a biased way as mentioned above, 
is a valuable advantage, which makes it easier to detect 
local density variations.

Relevant physical parameters for clusters
were determined in the following way. X-ray luminosities in the 0.1 to
2.4 keV energy band have been derived within a cluster radius of 
$r_{500}$ \footnote{$r_{500}$ is the radius where the average
mass density inside reaches a value of 500 times the critical density
of the Universe at the epoch of observation.}. To estimate the cluster
mass and temperature from the observed X-ray luminosity we use the 
scaling relations described in Pratt et al. (2009). These were
determined from a representative cluster sub-sample of our survey,
called {\sf REXCESS} (B\"ohringer et al. 2007). Since the radius 
$r_{500}$ is determined from the cluster mass, the calculation of 
X-ray luminosity inside $r_{500}$, cluster mass, and temperature
were performed iteratively, as described in B\"ohringer et al. (2013). 
The definitive identification of the clusters and the redshift 
measurements are described in Guzzo et al. (2009), Chon \& B\"ohringer 
(2012), and B\"ohringer et al. (2013).

The survey selection function was determined as a function
of the sky position with an angular resolution of one degree
and as a function of redshift. The selection function
takes all the systematics of the RASS exposure distribution, Galactic
absorption, the fiducial flux, and the detection count limit into account.
The interstellar hydrogen column density for these calculations is
taken from Dickey and Lockman (1990).
The selection function as a function of sky position and redshift was 
published for {\sf REFLEX II} in the online material of B\"ohringer et al. (2013)
and for  {\sf NORAS II} in  B\"ohringer et al. (2017a).

\section{Methods}

We studied the density distribution of clusters and of the underlying 
matter distribution as a function of redshift in different regions
of the sky. Because we used a flux-limited cluster sample with 
additional smaller sensitivity variations in regions of the sky with shorter 
exposures, we could not use the cluster number distribution directly
without taking the selection function into account. In the following
we used two different methods to achieve this.

In the first method we compared the observed cluster counts in
redshift bins with the expected counts. The expected counts were calculated
from the observed X-ray luminosity function convolved with
the survey selection function, which is given as a function of 
redshift and sky position.
For the luminosity function we took the best-fitting Schechter
function for the  {\sf REFLEX II} cluster survey from
B\"ohringer et al. (2014). The X-ray luminosity function for
{\sf NORAS II} is the same within the uncertainty limits (B\"ohringer 
et al. 2017). The  {\sf REFLEX II} luminosity function is shown in 
Fig. A1 of the Appendix and the parameters for the Schechter function are listed in 
Table A1, where we also give the parameters for the bracketing lower
and upper limit functions. We did not detect any significant evolution
in the X-ray luminosity function in the redshift range $z = 0 - 0.4$,
as shown and explained in detail in B\"ohringer et al. (2014). We therefore
assume this function to be constant over the distance range considered here.  
The relative density variations were then determined 
by the ratio of the observed to the expected number of galaxy clusters.

The second method was used to derive the unbinned cumulative mean 
density of clusters as a function of redshift. In this 
approach we attributed weights to each cluster to correct for
the spatially varying survey limits. The weights were calculated
from an integration of the luminosity function, $\phi(L_X)$, 
as follows:

\begin{equation}
w_i = {\int_{L_{X_0}}^{\infty} \phi(L) dL \over \int_{L_{X_i}}^{\infty} \phi(L) dL} ~~~, 
\end{equation}

where $L_{X_0}$ is the nominal lower limit of the sample 
and $L_{X_i}$ is the lower X-ray luminosity limit at the 
sky location and redshift of the cluster. We then determined 
the relative density distribution of the clusters by comparing 
the observed distribution of the clusters with
weights to the prediction of the cluster density for a volume complete sample
with a limiting luminosity of $L_{X_0}$. We used the same technique with
weights to produce maps of the projected density distribution of the clusters
in redshift slices. 

To infer the underlying matter distribution from the observed distribution 
of clusters, which is done in section 4.3, we assume 
that the cluster distribution is biased with
respect to that of the matter using the formalism of Tinker et al.
(2010). We verify this approach in Appendix B with
studies of cluster counts in cells in cosmological
numerical simulations. We find that the uncertainty in the prediction
of the matter density is roughly given by the Poisson error in the
cluster number counts.

To calculate the bias factor, which is independent of scale,
we used the formulas derived by Tinker et al. (2010) from
large N-body simulations. We calculated the bias as a function of cluster
mass for the adopted cosmological model~\footnote{The bias was calculated 
for a cosmological model with parameters of $\Omega_m = 0.282$ and
$\sigma_8 = 0.776$ which are consistent with the galaxy cluster observations
from our survey (e.g. B\"ohringer et al. 2014, 2017b).}. For easier application,
we fitted the result with a parameterised function of
the following form:

\begin{equation}
b(m) = A + B m + C m^2 + D m^{1/2}  + E m^{0.3}
,\end{equation}

with $A = 0.664$, $B = 0.1614$, $C = -1.23 \times 10^{-5}$, $D = 1.152$, 
and $E = 0.320$, where $m$ is the cluster mass, $M_{200}$, in units 
of $10^{14}~ h_{70}^{-1}$ M$_{\odot}$.   

The cluster mass was determined from the observed X-ray luminosity by means 
of the X-ray luminosity--mass relation
described in B\"ohringer et al. (2014), the same scaling relation used to
determine $r_{500}$ above. The mass estimate for individual clusters
has an estimated uncertainty of about 40\% (e.g. Pratt et al. 2009).
This translates into an uncertainty in the bias factor of not more than 5\%,
which we take into account in our modelling.

\section{Results}

\subsection{CLASSIX survey}

\begin{figure}
   \includegraphics[width=\columnwidth]{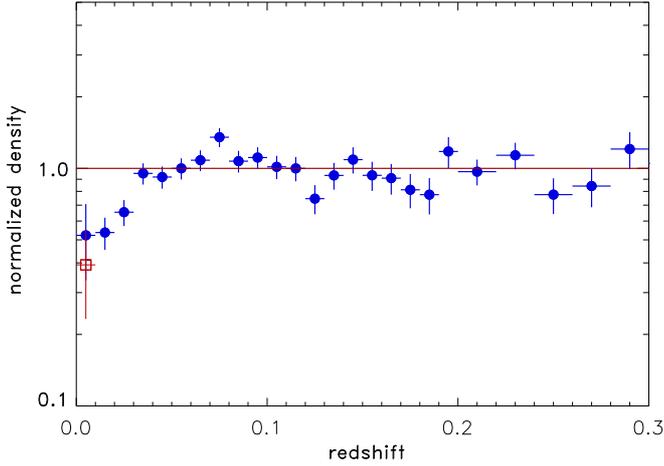}
\caption{Cluster density distribution as a function of 
redshift for the {\sf CLASSIX} galaxy clusters covering the sky 
at $|b_{II}| \ge 20^o$ for a minimum luminosity of 
$10^{42}$ erg s$^{-1}$ (0.1 - 2.4 keV). The density distribution has been
normalised by the expected cluster density based on the average luminosity
function as explained in section 3.
The open square shows the result if the region of the Virgo cluster
is excluded from the analysis.
}\label{fig1}
\end{figure}

In Fig.~\ref{fig1} we show the relative density distribution of 
the clusters for the entire {\sf CLASSIX} cluster sample with 
$L_x \ge 10^{42}$ erg s$^{-1}$ out to a redshift of $z = 0.3$,
excluding the zone of avoidance.
This distribution was constructed by dividing
the observed number of {\sf CLASSIX} clusters in different redshift bins 
by the prediction based on the best-fitting Schechter 
X-ray luminosity function and the 
{\sf CLASSIX} selection function. All the relative differential density
distributions of clusters shown in the following are constructed in
this fashion. Here, 211 clusters are involved in tracing the
density at $z \le 0.04$ and 1570 out to  $z = 0.3$.
While the overall cluster distribution 
is remarkably homogeneous, we note an underdensity of 
about 30 - 50\% at $z \le 0.03$  ($\sim 120$ Mpc). 

Because we are part of the Local Supercluster with the Virgo cluster
at its centre (where M87, M86, and M49 enter our catalogue as separate mass
halos) and since the X-ray emission of Virgo is partly blinding the region
behind the cluster, one could question if the sky region of the Virgo cluster 
should be included in our study. The open square in Fig.1 demonstrates 
that exclusion of the Virgo region has no effect 
on the further results of this paper.

Care needs to be taken in the interpretation of the local
underdensity observed in Fig.~\ref{fig1}. Since the region at very low
redshifts, which appears underdense, is traced mostly by 
objects with low  X-ray luminosity, which are only detected
in this region, there is some degeneracy in the determination of the 
X-ray luminosity function at the low-luminosity end 
and the relative cluster
density distribution in the nearby Universe. An overestimate
of the X-ray luminosity function at the low-luminosity end would
produce an artificial underdensity with the method applied here.

\begin{figure}
   \includegraphics[width=\columnwidth]{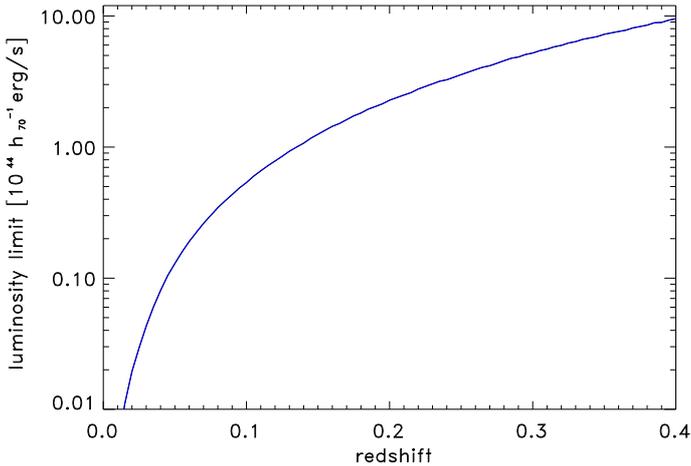}
\caption{Mean X-ray luminosity limit as a function of redshift for the
{\sf CLASSIX} survey.
}\label{fig2}
\end{figure}
 
\begin{figure}
   \includegraphics[width=\columnwidth]{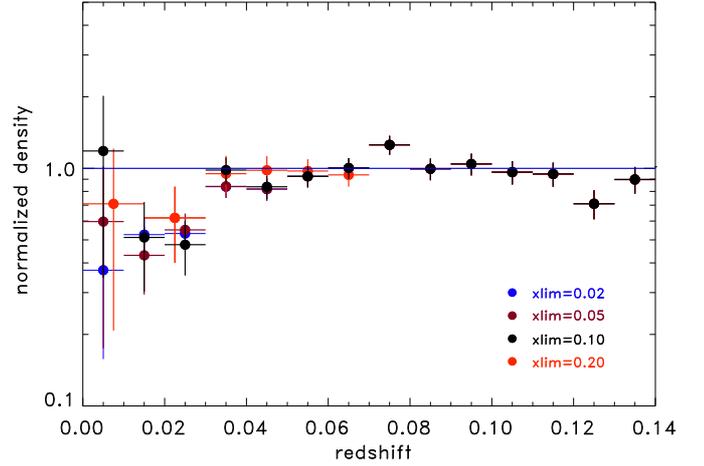}
\caption{ {\sf CLASSIX} galaxy cluster density distribution as a function of 
redshift for four different lower X-ray luminosity limits, given 
in the plot by the parameter $xlim$ in units of 
$10^{44}$ erg s$^{-1}$. All samples trace the same local density
deficit.
}\label{fig3}
\end{figure}

A way to break this ambiguity is to study a volume-limited sample
of clusters with a homogeneous lower X-ray luminosity limit over
a region that is larger than the observed underdensity. In Fig.~\ref{fig2}
we show the mean lower luminosity limit of the {\sf CLASSIX}
survey as a function of redshift. \footnote{The redshift limit is
independent of the adopted cosmological model because the luminosity
is determined from the flux with a cosmology-dependent luminosity distance,
while the redshift limit is in turn calculated from the limiting
luminosity using the inverse function of the same luminosity distance,
which cancels the dependence on cosmology.} 
We note that for example 
for an X-ray luminosity limit of $2 \times 10^{43}$ erg s$^{-1}$
we can sample the cluster density in a volume-limited way 
out to a redshift of $z = 0.062$, larger than the underdense region.
Therefore we constructed several cluster samples with
a range of lower limiting luminosities
($L_{x_0} = 0.02, 0.05, 0.1, 0.2 \times 10^{44}$ erg s$^{-1}$), 
which are volume limited out
to $ z =  0.032, 0.044, 0.062, 0.086$, respectively. 
The density distributions of these samples are shown in Fig.~\ref{fig3}. 
There is good agreement between the different samples and
they all trace a similar local underdensity. Therefore the observed
deficit cannot simply be the result of an inaccurately determined
X-ray luminosity function. We had shown a similar exercise
with the {\sf REFLEX II} survey in B\"ohringer et al. (2015)
with the same conclusion.

\subsection{Different hemispheres}

\begin{figure}
   \includegraphics[width=\columnwidth]{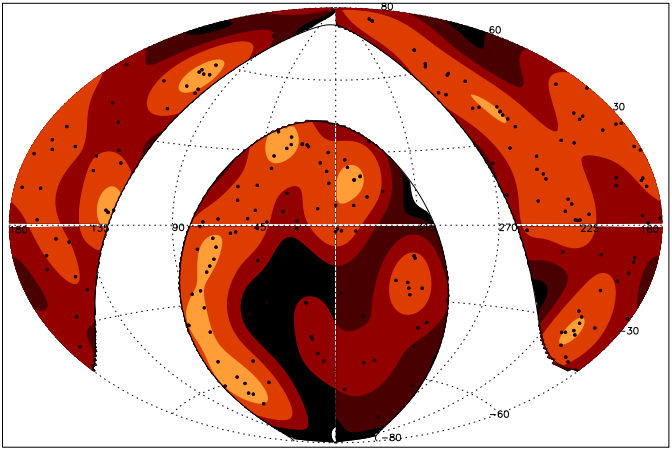}
\caption{Sky distribution of the clusters (black dots) and their surface density
in the {\sf CLASSIX} survey at $|b_{II}|\ge 20^o$ smoothed with a Gaussian filter
with $\sigma = 10^o$ in the redshift slice
$z = 0 - 0.04$ in equatorial coordinates. 
The colour coding for the density normalised to the mean is
orange: $> 2$, red: $1 - 2$, brown: $ 0.5 - 1$ , and dark brown/black: $< 0.5$.
}\label{fig4}
\end{figure}

Figure~\ref{fig4} shows the projected density distribution of 
the clusters in the redshift range $z = 0 - 0.04$. The 
colour-coded density 
distribution is that of the clusters with weights smoothed by a 
Gaussian filter with a $\sigma$-value of 10 degrees.
The density has
been normalised by the mean, so that the light(dark) regions show
overdensities (underdensities). We clearly note that the distribution
is not homogeneous, and so we do not expect to observe the same 
density deficit as noted in the mean radial profile in Fig.~\ref{fig1}  
in all sky directions. In the following
we therefore study how the local density distribution depends on the region in
the sky. 

\begin{figure}
   \includegraphics[width=\columnwidth]{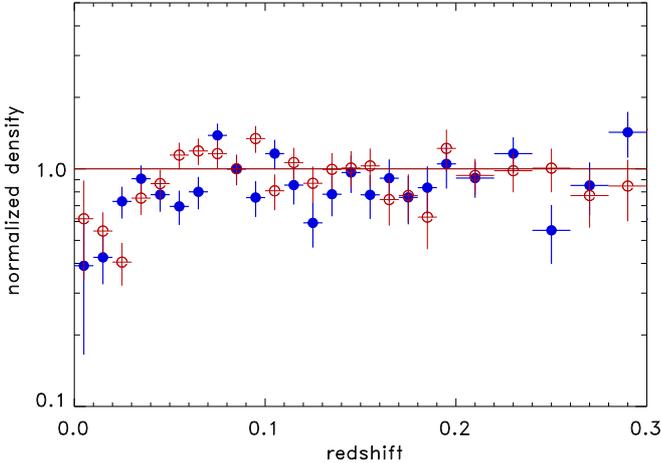}
\caption{Cluster density distribution as a function of 
redshift for the {\sf REFLEX II} survey in the southern sky (open red circles)
and the {\sf NORAS II} survey in the north (filled blue circles) 
at $|b_{II}| \ge 20^o$, 
for a minimum luminosity of $10^{42}$ erg s$^{-1}$ (0.1 - 2.4 keV).
The density distribution has been
normalised by the expected cluster density based on the average luminosity
function as explained in section 3. 
}\label{fig5}
\end{figure}

In Fig.~\ref{fig5} we show the cluster density distribution in the northern sky 
({\sf NORAS II}) and southern sky ({\sf REFLEX II})
at $|b_{II}| \ge 20^o$. Here the
{\sf REFLEX II} survey extends towards the north to a declination
of $+ 2.5^o$, overlapping slightly with the {\sf NORAS II} survey.
While the extent of the local deficit in the north reaches
a redshift of about $z \sim  0.02$ ($\sim 85$ Mpc), that in the 
southern sky stretches out to about
$z \sim 0.04$ ($\sim 170$ Mpc). At larger reshifts the 
distribution is again relatively homogeneous.

\begin{figure}
   \includegraphics[width=\columnwidth]{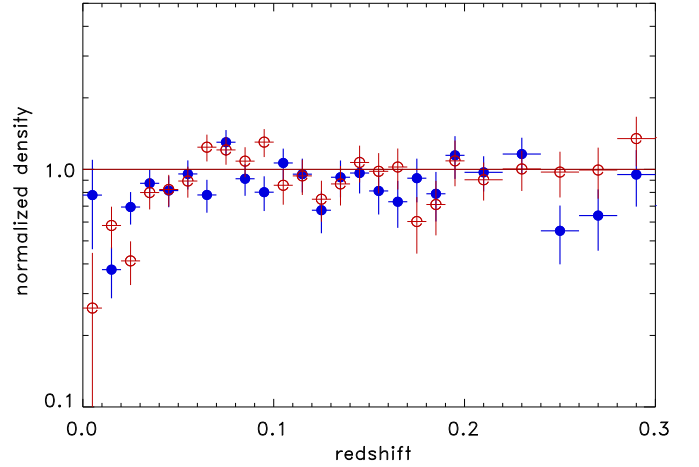}
\caption{Cluster density distribution as a function of 
redshift in the northern Galactic cap (filled blue circles) and
southern Galactic cap (open red circles) at $|b_{II}| \ge 20^o$, 
for a minimum luminosity of $10^{42}$ erg s$^{-1}$ (0.1 - 2.4 keV).
The density distribution has been
normalised by the expected cluster density based on the average luminosity
function as explained in section 3.
}\label{fig6}
\end{figure}

The density distributions in the northern and southern Galactic hemisphere 
(at $|b_{II}| \ge 20^o$) are compared in Fig.~\ref{fig6}. 
The underdensity is less pronounced in the northern Galactic cap,
with a deficit of about 35\% at $z \le 0.03$ compared 
to about 47\% in the south. 
In the south the density deficit stretches out
to about 130 Mpc. 

\begin{figure}
   \includegraphics[width=\columnwidth]{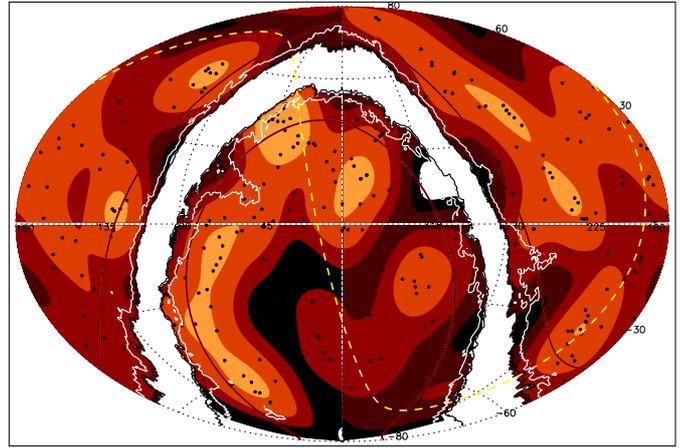}
\caption{Sky distribution of the clusters and their surface density
in the {\sf CLASSIX} survey with the extension into the
zone of avoidance in equatorial coordinates. 
The survey is bounded by an interstellar
hydrogen column density limit of $n_H \le 2.5 \times 10^{21}$ cm$^{-2}$.
The white contours show the hydrogen column density boundary of 
$n_H = 1.5 \times 10^{21}$ cm$^{-2}$. The red lines indicate the Galactic
latitudes of $|b_{II}| = \pm 20^o$. The yellow dashed line marks the 
Supergalactic plane and the colour coding is the same as in  Fig.~\ref{fig4}.
}\label{fig7}
\end{figure}

To see if the local cluster density in the sky outside the band of the Galaxy
may be compensated by an overdensity in the zone of avoidance, we looked
into our incomplete survey of this region. Figure~\ref{fig7}  shows the cluster
distribution across the sky, now with part of the region 
of the zone of avoidance, which is  
covered by our survey. The survey area is limited by an interstellar hydrogen
column density of $n_H \le 2.5 \times 10^{21}$ cm$^{-2}$. We also show 
the region with a limit of
$n_H \le 1.5 \times 10^{21}$ cm$^{-2}$ bounded by white contours that
was explored alternatively.
The figure shows in addition the Galactic band  
($|b_{II}| \ge 20^o$, black lines) and the location of the Supergalctic plane.

\begin{figure}
   \includegraphics[width=\columnwidth]{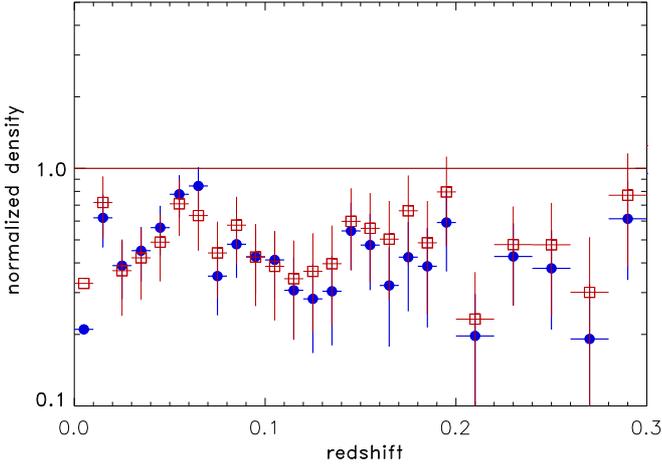}
\caption{Cluster density distribution as a function of redshift 
in the zone of avoidance at $|b_{II}|  < 20^o$. Filled symbols are for
the region with a Galactic hydrogen column density of $n_H < 2.5 \times 10^{21}$
cm$^{-2}$ and open symbols for $n_H < 1.5 \times 10^{21}$ cm$^{-2}$. The
cluster sample in these region is not complete and therefore the data
provide a lower limit.
The density distribution has been
normalised by the expected cluster density based on the average luminosity
function as explained in section 3.
}\label{fig8}
\end{figure}

The zone of avoidance does not show any large local overdense regions
as displayed in Fig.~\ref{fig8}. 
We roughly expect that our survey has a completeness of about 60 - 70\%
including the incomplete spectroscopy follow-up. This 
incompleteness is at least
partly responsible for the lower value of the mean density in the figure. 
We note that so far we
have no evidence of an overdensity of clusters behind the band
of the Galaxy.

\subsection{Cumulative densities}

To probe the density distribution on a finer scale we now use the second 
method described in section 3 to show the unbinned cumulative density of 
the clusters, that is the mean density inside a certain distance taken 
at the redshift of each cluster. For this we sum the clusters multiplied 
with their weights and compare this with the number of clusters we would expect 
in a volume-limited sample out to the same distance with the adopted lower
luminosity limit of the analysis.  

\begin{figure}
   \includegraphics[width=\columnwidth]{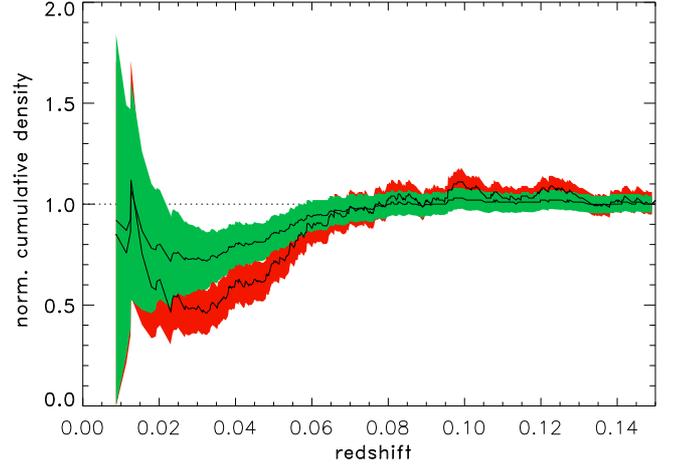}
\caption{Cumulative density distribution  of {\sf REFLEX II} clusters as
a function of redshift normalised to the mean for a lower X-ray luminosity
limit of $L_{X_0} = 5 \times 10^{42}$ erg s$^{-1}$ (lower curve
with red uncertainty limits). The upper curve with green uncertainty limits
shows the inferred dark matter distribution after correcting for the
cluster bias.
}\label{fig9}
\end{figure}

In Fig.~\ref{fig9} we show the cumulative density distribution of the 
{\sf REFLEX II} clusters in the southern sky normalised to the 
mean density. To minimise the influence of the low-luminosity end
of the X-ray luminosity function we used a lower luminosity
limit of $L_{X_0} = 5 \times 10^{42}$ erg s$^{-1}$ here.
The plot shows that the underdensity reaches out to about 
$z \sim 0.04$ as in the differential plot above, but despite the
local overdensity at the boundary of the underdense region, the cumulative
mean density is only recovered at $z \sim 0.06$. We also show the uncertainty limits
as a red region, which takes into
account the uncertainty of the X-ray luminosity function (Fig. A1)
used for the normalisation and the Poisson error of the cluster number 
counts.

Figure~\ref{fig9} also shows the inferred underlying matter distribution traced
by the clusters. We derive this by accounting for the fact that clusters
follow the matter distribution in a biased way. We corrected for
the bias in the way described in section 3 and included an additional
uncertainty in the estimated bias factor due to uncertainties in the 
mass of galaxy clusters.
We note a mean matter underdensity of about  $-27 \pm 15\%$
out to $z \sim 0.033$ ($\sim 140$ Mpc) and of about  $-20 \pm 10\%$ out
to $z \sim 0.045$ ($\sim 190$ Mpc).

\begin{figure}
   \includegraphics[width=\columnwidth]{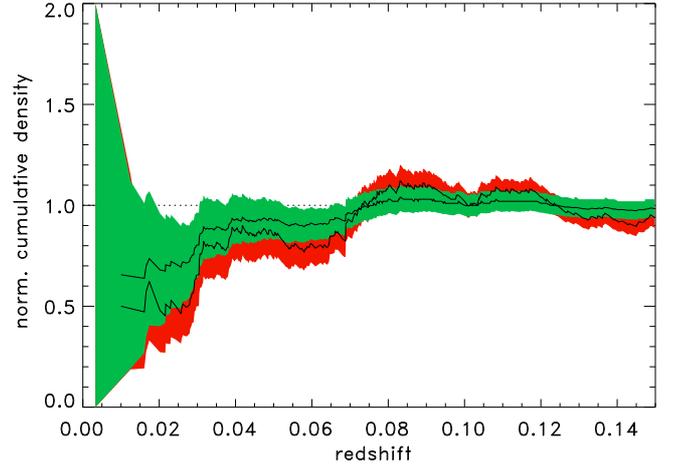}
\caption{Cumulative density distribution  of {\sf NORAS II} clusters as
a function of redshift normalised to the mean for a lower X-ray luminosity
limit of $L_{X_0} = 5 \times 10^{42}$ erg s$^{-1}$ (lower curve
with red uncertainty limits). The upper curve with green uncertainty limits
shows the inferred dark matter distribution after correcting for the
cluster bias.
}\label{fig10}
\end{figure}

Figure~\ref{fig10} shows in a similar way the cumulative cluster density 
distribution in the northern sky at $|b_{II}| \ge 20^o$.  The local underdensity
is deeper ($-50\% \pm 20\%$), but at this depth it only extends 
to about 90 Mpc. In the cumulative density we see, after a sharp density increase,
a slow recovery 
of the mean density which is reached at $z \sim  0.07$.
For a mean matter underdensity of $-30\% \pm 15\%$ the extent of the region is
about 100 Mpc ($z \sim 0.024$) and for $-20\% \pm 10\%$ it reaches
130 Mpc ($z \sim 0.03$).  

\begin{figure}
   \includegraphics[width=\columnwidth]{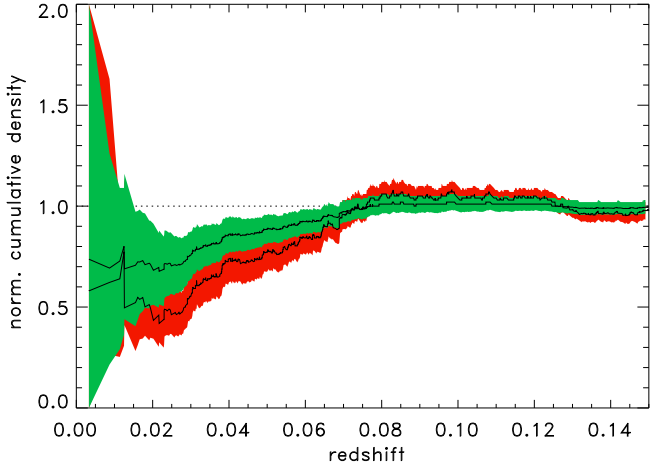}
\caption{Cumulative density distribution  of {\sf CLASSIX} clusters as
a function of redshift normalised to the mean for a lower X-ray luminosity
limit of $L_{X_0} = 5 \times 10^{42}$ erg s$^{-1}$ (lower curve
with red uncertainty limits). The upper curve with green uncertainty limits
shows the inferred dark matter distribution after correcting for the
cluster bias. 
}\label{fig11}
\end{figure}

In Fig.~\ref{fig11} we show the same plot for the entire 
{\sf CLASSIX} survey at $|b_{II}| \ge 20^o$. The results show
approximately a mean behaviour of that of the two hemispheres.
For a mean matter underdensity of $-30\% \pm 15\%$ the extent of the region is
about 100 Mpc ($z \sim 0.0235$) and for $-20\% \pm 10\%$ it reaches
about 140 Mpc ($z \sim 0.033$).

\subsection{Particular sky regions}

\begin{figure}
   \includegraphics[width=\columnwidth]{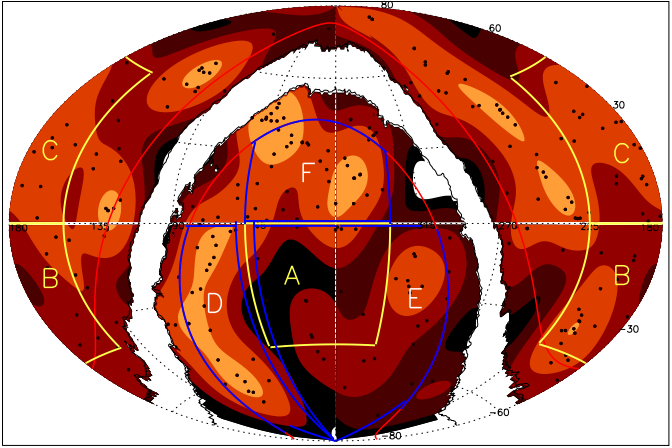}
\caption{Sky distribution of the clusters and their surface density
in the extended {\sf CLASSIX} survey in equatorial coordinates. 
Particular regions marked and
labelled in the figure are explained in the text. The red lines mark
the Galactic latitudes $|b_{II}| \pm 20^o$ and the displayed survey region
is limited by an interstellar hydrogen column density value of 
 $n_H \le 2.5 \times 10^{21}$ cm$^{-2}$. The yellow lines mark
the boundaries of regions A to C and the blue lines those of regions
D to F.
}\label{fig12}
\end{figure}

We also inspected the density distribution in smaller regions of the sky.
However, the smaller number statistics increases
the uncertainties.  We have already analysed two particular regions in our earlier study of the southern sky, where we can compare
our cluster distribution to observations of the galaxy density
distribution from Whitbourn \& Shanks (2014). These are the sky areas 
labelled A and B in Fig.~\ref{fig12}. We found remarkably 
good agreement between galaxy density and cluster density in these 
sky areas (B\"ohringer et al. 2015, Figs. 8 and 9). 

\begin{figure}
\includegraphics[width=\columnwidth]{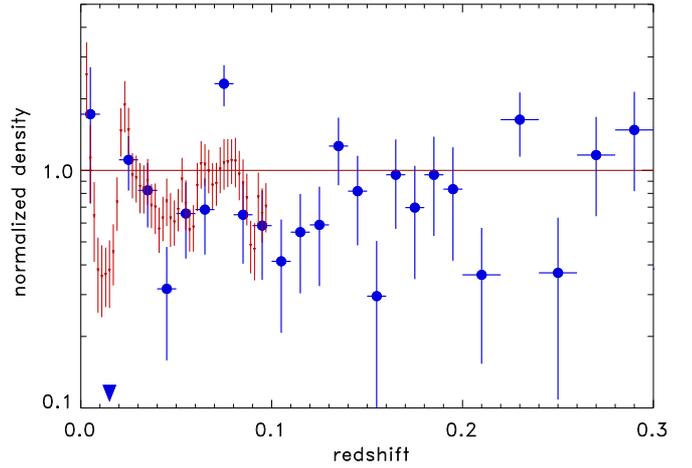}
\caption{Density distribution of {\sf CLASSIX} clusters as a 
function of redshift in the region labelled C in Fig.~\ref{fig12}. 
We find no cluster in the second redshift bin marked 
by a downward pointing triangle. The galaxy
distribution (Whitbourn \& Shanks 2014) in the same area
is shown by smaller red points with error bars. There 
is good agreement between both density distributions.
}
\label{fig13}
\end{figure}

The third region studied by Whitbourn \& Shanks (2014) in the equatorial
northern part of the north Galactic cap, region C in Fig.~\ref{fig12},
is explored in  Fig.~\ref{fig13}. There is no underdense region in this
area, except for the redshift bin $z = 0.01 - 0.02$ where we find no
cluster above our flux limit. 
The galaxy distribution follows that of the clusters closely and
in the redshift bin where we detect no cluster, we also note
a pronounced underdensity in the distribution of galaxies.
The fact that galaxies and clusters show
approximately the same density distribution provides further strong support 
that the {\sf CLASSIX} clusters are fair tracers of the underlying
matter distribution.

\begin{figure}
\includegraphics[width=\columnwidth]{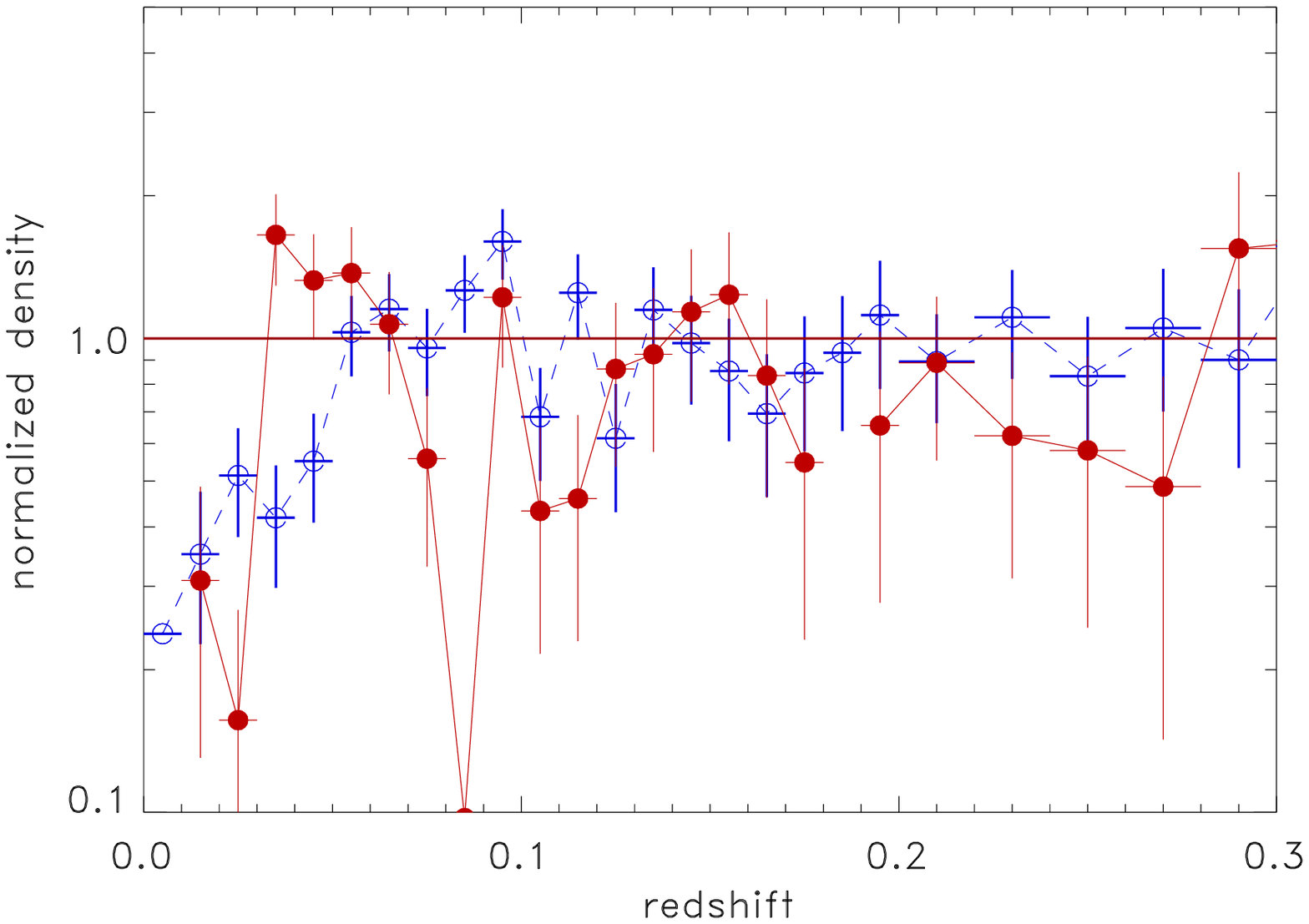}
\includegraphics[width=\columnwidth]{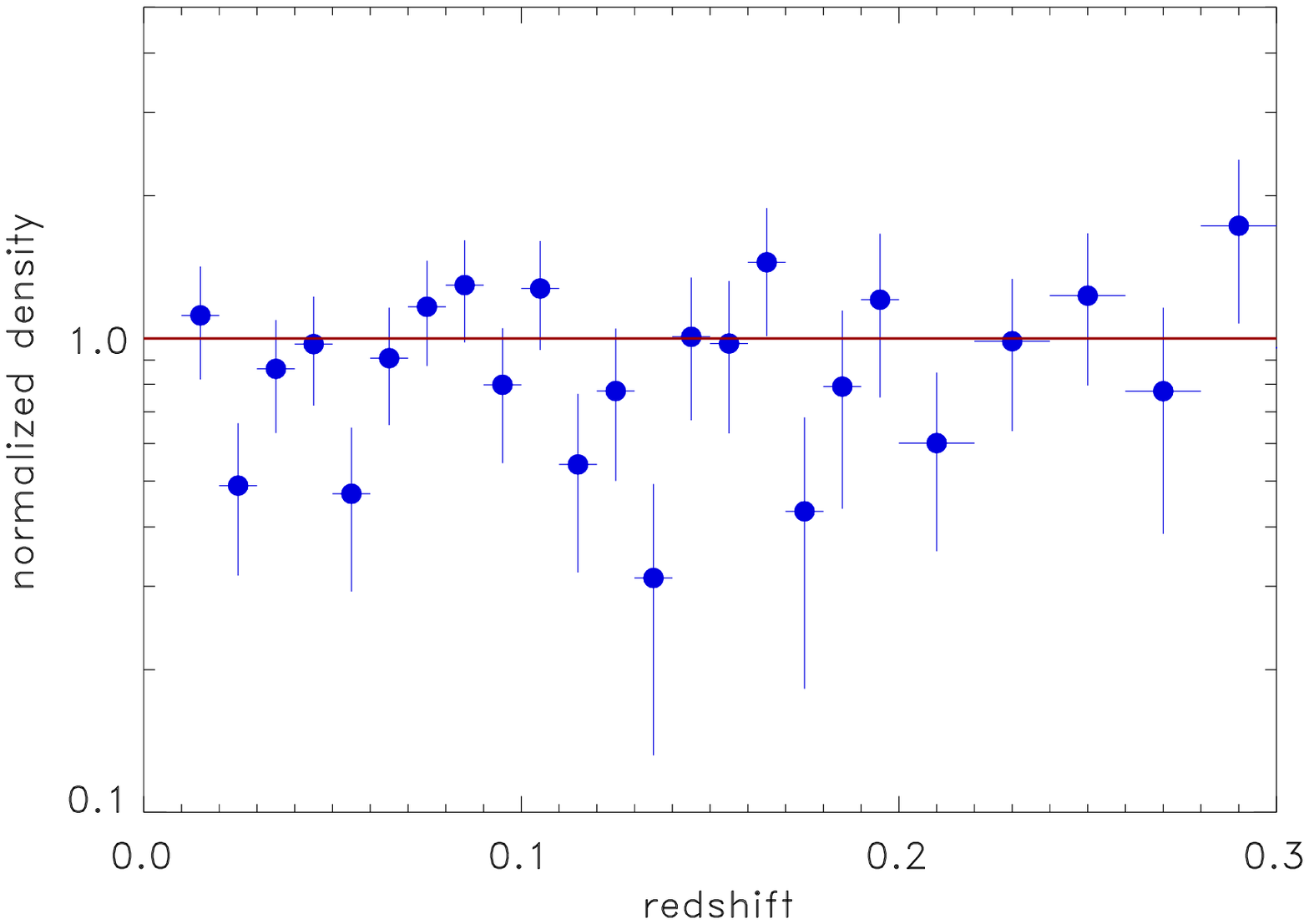}
\caption{{\bf Top}: Density distribution of {\sf CLASSIX} clusters as a 
function of redshift in the high-density region in the southern sky, D 
(red filled circles), and  the low-density region, E (blue open circles). 
{\bf Bottom:}  Density distribution of {\sf CLASSIX} clusters as a 
function of redshift in the northern high-density region, F. 
This seems to be one of the densest regions at $z \le 0.04$. 
}
\label{fig14}
\end{figure}

To further explore the variance in the cluster density distribution 
in different celestial regions, we selected a few sky areas that
show a particularly high or low density in Fig.~\ref{fig12}.
The regions labelled D and E in the figure (with right ascension and 
declination ranges of $RA = 55 - 115^o$, $DEC \le 0^o$
and $RA  \le 45^o ,\ge 270^o$, $DEC \le 0^o$, respectively
and $|b_{II}| \ge 20^o$) are shown in the top
panel of Fig.~\ref{fig14}. While the denser region D shows a nearby cluster
deficit, the region is characterised by an overdensity at redshift
$z = 0.03 - 0.04$.
The region E around the south Galactic pole shows a particularly pronounced
underdensity. In the northern sky we have the region F 
($RA  \le 50^o ,\ge 330^o$, $DEC \ge 0^o$ and $|b_{II}| \ge 20^o$) 
which shows a higher-than-average
density in Fig.~\ref{fig12}. The density distribution  
of F shown in the bottom panel of Fig.~\ref{fig14} is
mostly overdense and does not contribute to the overall local
underdensity at all. In summary we note that the underdensity
in the local Universe has a complex structure and a homogeneous
spherical void would be a rather crude representation of its geometry.

\section{Discussion}

Combining the results from section 4.3, we infer from the observed
cumulative cluster density distribution a local underdensity with a 
deficit of $-0.3 \pm 0.15$ extending about 100 Mpc to the north and 
of $-0.27 \pm 0.15$ extending about 140 Mpc to the south. 
This underdensity is bounded by well-known
superclusters. In the northern sky it ends at the Great Wall,
while in the south its boundary is at 
the Shapley supercluster and two further superclusters, 
RXSCJ0338-5414 (at $z = 0.0603$) and RXSCJ0624-5319 (at $z = 0.0520$), 
identified by Chon et al. (2013) in our 
survey. These superstructures seem to terminate the underdensity. 
Among the superclusters in the local Universe, the Shapley
supercluster is by far the most prominent structure (e.g. Sheth 
\& Diaferio 2011, Chon et al. 2015). Therefore, one way to put the 
observation of the local underdensity into perspective is to
note that we do not live near one of the prominent superstructures.
The Local Supercluster (e.g. de Vaucouleurs 1959) is not one of the massive
superclusters. Therefore, the large-scale mean matter density of the 
Universe seems to be fairly sampled only when the volume is large 
enough to include also the very massive superstructures.

An important question to ask is how likely it is
to find the observed extended underdensity in a Universe
described by the concordance $\Lambda$CDM cosmological model.
To answer the question we adopted an approximate description
of the observed underdensity by a spherical region with a radius 
of about 100 Mpc radius, as found for the {\sf CLASSIX} survey
corresponding to an underdensity of $-0.3 \pm 0.15$. In linear 
theory we can calculate the probability of finding such a region 
from the variance of the matter density
distribution filtered by a top-hat filter with the given radius.
To infer the linear density from the observed underdensity we have
to correct for the extra expansion of the region in the non-linear
evolution, yielding a Lagrangian radius of $92 \pm 4$ Mpc. 
Using the power spectrum for the $\Lambda$CDM cosmological 
model that best fits our cluster data
(B\"ohringer et al. 2014), we can calculate the RMS fluctuation
amplitude for this scale. We applied {\sf CAMB} (Lewis et al. 2000)
\footnote{ CAMB is publicly available from
http://www.camb.info/CAMBsubmit.html}
to obtain the matter power spectrum. For the RMS amplitude
we obtained values of $\sigma = 0.115 \pm 0.005$. Therefore, an
underdensity of the above given amplitude corresponds to a
$1.3 - 3.8\sigma$ deviation from the mean density. 
For the lower limiting value the 
probability of finding such an underdensity is therefore about 
10\%, a possibility that cannot easily be ruled out.
If we look alternatively at the region which has a mean
underdensity of  $-0.2 \pm 0.1$ and a radial extent of about 
140 Mpc, we obtain the following values: the radius in linear
approximation is $\sim 132 \pm 4$ Mpc and the RMS 
fluctuation amplitude is $\sigma = 0.075 \pm 0.003$,
corresponding to a $1.4 - 3.9\sigma$ excursion. Considering these
results, it seems more likely that the true values for the matter 
density deficit are close to our lower uncertainty limits.

Several works studied the probability of a 
local matter underdensity with similar results (e.g. 
Yu 2013, Wojtak et al. 2014, Odderskov et al. 2017, Wu \& Huterer 2017,
Fleury et al. 2017).
Among these studies, it is interesting to mention the result of
Wojtak et al (2014), who discussed conditional 
probabilities. In the case where one wishes to know the probability of
the density distribution observed from a random point in space,
the probability of finding oneself in a void  is slightly higher,
since underdense regions occupy more space in non-comoving units 
than overdense regions. However, if one applies the condition that 
the observer is located in a dark matter halo with a mass of
about $10^{13}$ M$_{\odot}$, which may describe the properties
of the Local Galaxy Group, the chance of being located in an 
overdense region is slightly higher. Despite the fact that 
the second case should be a better representation of the real
situation, we seem to find ourselves in an underdense area.

Another consideration is the chance that the sky region hidden
behind the Milky Way could compensate the deficit seen in the
{\sf CLASSIX} survey. If we take the entire region at 
$|b_{II}| < 20^o$, which is roughly half the area of {\sf CLASSIX},
we would need a matter overdensity of about 60\% out to a radius of
100 Mpc. Calculating the probability for this to happen in
a $\Lambda$CDM cosmological model in a similar way as above, we
find a $\sigma$-value for the probability of $3.8\sigma$, hence much
less likely than the value for a 30\% underdensity in the
{\sf CLASSIX} area ($2.6\sigma$). According to Tully et al. 
(2019) the ``Local Void'', one of the largest underdense 
structures nearby, is mostly hidden by the zone of avoidance.
Since the analysis by Tully et al. is based on peculiar velocities,
their method is also sensitive to structures not directly observed.
Thus they can in principle obtain a more complete picture (in
a smaller redshift region) than what we can presently map with the
cluster distribution. Therefore, the existence of the Local Void
in the hidden region behind the band of the Milky Way makes
it even more unlikely that the zone of avoidance can compensate
the observed local matter deficit.   

In a recent study, Jasche \& Lavaux (2019) used the 
2M++ galaxy sample compiled by Lavaux \& Hudson (2011)
based on the 2MASS Galaxy Redshift Survey (Huchra et al. 2012)
for a reconstruction of the matter density distribution in the 
local Universe with a Bayesian modelling technique including
the use of N-body simulations for cosmic structure evolution. 
One of their results provides radial matter density 
distributions averaged in shells around our location
presented in their Fig. 10. The density profile for the whole
sky, shown in the left panel of that figure, features more underdense
than overdense regions out to a redshift of about 150 h$^{-1}$ Mpc.
If this differential density profile is integrated, the cumulative
profile shows a mean underdensity of about 10 - 20\% inside a 
radius of about 85 Mpc. The result is qualitatively very
similar to ours, but the extent and the amplitude of the underdensity are 
somewhat smaller. With our large uncertainties the two results could 
be considered marginally consistent. There are however two possible 
reasons for the difference. First, the Bayesian method includes the
$\Lambda$CDM model with approximately Planck Mission constraints for
the cosmological parameters as a prior, which means that the consistency
with this model is also driving the results. Second, the galaxy 
sample is limited to
redshifts below $z = 0.06$ to $0.08$ (their Fig. 2). Their reference of the 
large-scale mean density therefore comes from a smaller volume than ours, and
so it may be difficult to detect an underdensity with a larger
extent than what they find. In light of these considerations
we interpret both results as consistent.
Figure 10 of Jasche \& Lavaux (2019) also shows the radial profiles
for two survey regions of Whitbourn \& Shanks (2014) labelled A and B
above. In both regions we observe a similar density structure as
outlined by the clusters and galaxies.

If the density of the local Universe is less than the mean density, the Hubble 
constant measured within this volume is larger than that
found at larger scales. In Appendix C we calculate how
the Hubble constant depends on the density. For a deficit
of  $-0.3 \pm 0.15$ we find a value for $H_0$ which is higher
by $5.5^{+2.1}_{-2.8} \%$, and for$-0.2 \pm 0.1$ the increase of $H_0$ 
would be $3.5^{+1.9}_{-1.8} \%$. We note
that these values of the Hubble constant refer
to the volume of the underdensity. Most local measurements of $H_0$
cover a larger volume, for example those of Riess et al. (2019), where the
described effects are diluted in the average result. 
The quoted changes of $H_0$ apply,
however, to measurements inside the underdensity, like all studies
based on peculiar motions; for example those of Tully et al. (2016, 2019b),
which imply a value of $H_0$ of about 75 km s$^{-1}$ Mpc$^{-1}$.

Determining the Hubble constant in the redshift range $z = 0.018 - 0.85$
 using a distance calibration from the analysis of Baryonic acoustic
oscillations (BAOs) in the Dark Energy Survey, independent of local distance
calibrators, Macaulay et al. (2019) find a Hubble constant of 
$H_0 = 67.8 \pm 1.3$ km s$^{-1}$ Mpc$^{-1}$. The good agreement
with the results from the Planck mission is not surprising, since
both analyses rely on the sound horizon as a calibration standard.

In a recent update on their work, Shanks et al. 2019b modelled their
data on the galaxy density distribution with a self-consistent outflow
model, finding that the Hubble constant would be increased 
by about 2 - 4\% inside a region with a radius of about $150 h^{-1}$ Mpc.
Lukovic et al. (2019) explore the evidence 
of a local void with SN data from the joint light curve analysis (JLA, 
Betoule et al. 2014) and Pantheon sample (Scolnic et al. 2018) using
a Lema\^itre-Tolman-Bondi cosmological model. Lukovic et al. find
constraints on a local underdensity with a size of $z \le 0.039^{+0.062}_{-0.018}$
and a density contrast of $\delta\rho/\rho = -7.5^{+12.9}_{-11.0}\%$ for JLA
as well as $z \le 0.070^{+0.023}_{-0.031}$ and $\delta\rho/\rho = -7.4^{+10.5}_{-7.0}\%$
for the Pantheon sample. The results are therefore consistent with homogeneity, 
but also within 1$\sigma$ errors with a local underdensity as 
found by Whitbourn \& Shanks
(2014) and with our findings. These latter authors also study the implications for the
galaxy distribution of Keenan et al. (2013), for which they obtain the
constraints of a void size of  $z \le 0.079^{+0.012}_{-0.012}$ with an underdensity
of $\delta\rho/\rho = -43.8^{+6.0}_{-6.1}\%$. This result is inconsistent
with the SN data however, excluding one critical data point
out of ten relaxes this discrepancy and also makes these findings
more similar to our results. More stringent constraints were obtained by
Kenworthy et al. (2019) with the Pantheon sample combined with the 
Foundation survey and the Carnegie Supernova Project, excluding a local 
underdensity of $\sim 100$ Mpc in size with a density contrast of $\delta\rho/\rho > 27\%$
at 5$\sigma$, which does not rule out our results closer to their lower limits.
In summary, the SN data are not in conflict with our findings.

\section{Summary and conclusion}

We find a significant local underdensity at redshifts 
$z \le  0.03 - 0.04$ in the distribution of galaxy
clusters, compared to the mean cluster
density over a large volume observed out to $z = 0.3$
(excluding the zone of avoidance, with $|b_{II}|  \le 20^o$) . It is well
known that clusters trace the density distribution of matter on large scales
in a statistical sense, and we have shown here (Appendix B) that
there is a tight correlation for the cluster density and matter density in 
cells of numerical simulations. We have also shown that this underdensity is traced
by several subsamples of our cluster catalogue, including for example
only the more X-ray luminous clusters. Therefore, we are sure that this is not
an effect of missing clusters in our survey and we have strong
evidence that this underdensity is real.

We studied the likelihood of finding such an underdensity in a universe
described by a concordance $\Lambda$CDM cosmological model 
\footnote{with cosmological 
parameters that are consistent with the statistics of the galaxy
cluster population and most other measurements of the local large-scale
structure} and found probabilities that are relatively small. But for underdensity 
amplitudes close to our lower uncertainty boundary, probabilities of 
$\sim 10\%$ are still large enough that such a case cannot easily be ruled
out for statistical reasons. 

As discussed in previous studies (see references in the introduction) a
local matter underdensity has consequences for the Hubble constant
measured with precision distance estimators in the low-redshift 
Universe. One of the currently heavily discussed problems of cosmological
measurements is the discrepancy in the Hubble constant inferred from
the analysis of the cosmic microwave background anisotropies observed
by Planck with a value of $67.4~ (\pm 0.5)$ km s$^{-1}$ Mpc$^{-1}$ 
(Planck Collaboration 2016, 2018) and
the values found from local estimators 
with a value of about $74.0~ (\pm 1.4)$ (e.g. Riess et al., 2019).
This is a  difference of about 9.6\%, much larger than the combined error.
Our finding can at least explain part of the difference. But the 
discrepancy is larger than what could plausibly be accommodated by
our observations. For most measurements of $H_0$ from SNe 
the volume of reliable measurements is larger than the underdensity
and the effect is further diluted.
Therefore, one has to look in addition for other reasons
for this discrepancy. There could well be further systematic effects 
which may have been overlooked or have been underestimated so far. On 
the other hand there is a growing number of publications which discuss physical
effects causing this difference in the Hubble constant (e.g.
Di Valentino et al. 2018, D'Eramo et al. 2018, Poulin et al. 2019,
Pandey et al. 2019, Vattis et al. 2019, Agrawal et al. 2019, Desmond
et al. 2019).

What remains important in any case is that the observations
of a local underdensity, for which we provide well-founded evidence,
have to be taken into account. Another important point of our findings
is that the underdensity is not seen in all regions of
the sky and therefore these variations across the sky need to be 
taken into account for precise cosmological calculations. 
So far only a few studies based on
the galaxy distribution support our conclusions (e.g. Keenan et al. 
2013, Whitbourn \& Shanks 2014), because a lot of work tracing the matter
distribution with galaxies does not extend as far as the size of the local 
underdensity. However, with the growing size and increased precision of 
ongoing and planned galaxy surveys we hope to  soon see firm confirmation of our 
observations from galaxy studies.     

\begin{acknowledgements}
We like to thank the anonymous referee for constructive comments.
We acknowledge informative discussions with Tom Shanks and useful comments
by Adam Riess, D'Arcy Kenworthy, Vladimir Lucovic, and Edward Macaulay.   
G.C. acknowledges support by the DLR under grant no. 50 OR 1905.
\end{acknowledgements}

\appendix
\section{X-ray luminosity function}

\begin{figure}
   \includegraphics[width=\columnwidth]{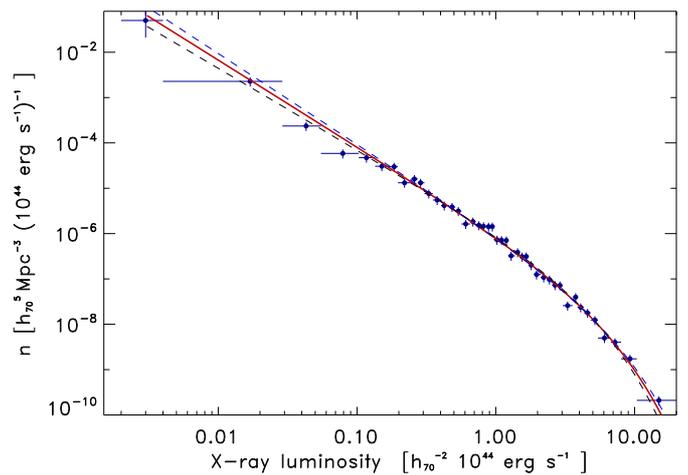}
\caption{{\sf REFLEX II} X-ray luminosity function for the redshift range
z = 0 - 0.4. We also show the best-fitting Schechter function
 and the uncertainty limits of the fit (B\"ohringer et al. 2014).
}\label{fig_A1}
\end{figure}

   \begin{table}
      \caption{Best-fitting parameters for a Schechter function describing
       the {\sf REFLEX II} X-ray luminosity function. For the description 
       of the parameters of the Schechter function see Eq. A.1.
        $L_X^{\ast}$ has units of $10^{44}$ erg s$^{-1}$ in the 0.1 - 2.4
        keV band and $n_0$ units of $h_{70}^5$ Mpc$^{-3}$ ($10^{44}$ erg s$^{-1}$)$^{-1}$.}
         \label{Tempx}
      \[
         \begin{array}{llll}
            \hline
            \noalign{\smallskip}
  ~~L_x-{\rm range}& ~~~~~~\alpha  & ~~~~~~L_X^{\ast} & ~~~~~~n_0 \\
            \noalign{\smallskip}
            \hline
            \noalign{\smallskip}
{\rm best}    & 1.92  & 3.95 & 2.83\cdot 10^{-7}  \\
{\rm low}    & 1.8  & 3.2 & 4.4\cdot 10^{-7}  \\
{\rm high}    & 2.0  & 4.7 & 2.0\cdot 10^{-7} \\
            \noalign{\smallskip}
            \hline
            \noalign{\smallskip}
         \end{array}
      \]
\label{tab1}
   \end{table}

The X-ray luminosity function of the clusters of our survey was
determined for the southern part ({\sf REFLEX II}) in 
B\"ohringer et al. (2014). We use this result here in its
parametric form, a Schechter function defined as

\begin{equation}
{n(L_X)~dL_X}~ =~ n_0~ \left( {L_X \over L_X^{\ast}}\right)^{-\alpha}
exp\left(- {L_X \over L_X^{\ast}}\right)  {dL_X \over L_X^{\ast}} ~~~.
\end{equation}

The {\sf REFLEX II}  X-ray luminosity function and the Schechter function fit
is shown in Fig.~\ref{fig_A1} and the parameters for the fitted
function are given in Table A.1 (B\"ohringer et al. (2014).
In addition to the best-fitting function we also use two bracketing 
functions, also given in the figure and the table, which capture the uncertainty 
in the fit of the Schechter function.
In our study in B\"ohringer et al. (2014) we found no significant evolution
of the X-ray luminosity function of the {\sf REFLEX II} clusters
in the redshift interval $z = 0$ to $0.4$. Therefore we assume this
function to be constant in the volume studied here. The X-ray luminosity
function determined from the {\sf NORAS II} survey agrees with that of
{\sf REFLEX II} within their uncertainties (B\"ohringer et al. 2017).

\section{Galaxy clusters tracing the matter distribution}

To investigate how well galaxy clusters trace the matter distribution
we used the Millennium simulations (Springel et al. 2005). While it 
is well known that clusters provide a biased account of the fluctuations
in the matter density distribution in a statistical analysis such as
the two-point-correlation function or the power spectrum, we 
tested here how well the cluster density correlates with the matter 
density in individual patches of the Universe. We therefore compared
cluster counts in cells to the mean matter density in the cells 
in the Millennium simulations.

The Millennium simulations are dark-matter-only simulations, which
is sufficient for our purpose, since we are looking at very large 
scales of tens of megaparsecs where baryonic effects play no significant
role. The cosmological parameters used for the Millennium study 
($\Omega_m = 0.25$, $\sigma_8 = 0.9$, and $H_0 = 73$ km s$^{-1}$ 
Mpc$^{-1}$) are different from our preferred cosmology.
Thus the bias is slightly different. However, here we are
not interested in calibrating the biasing relation, but we
want to demonstrate the method of tracing the matter distribution in
spatial patches and to study its uncertainty. For this purpose the 
difference in the cosmological parameters is not important.

The Millennium simulation has a box size of $500~ h_{100}^{-1}$ Mpc.
We selected clusters with a lower mass limit of $0.5 \times 10^{14}$ M$_{\odot}$
finding 5283 such systems in the simulation. We performed two
studies: one with a box size of $89.3~ h_{70}^{-1}$ Mpc and
one with $178.6~ h_{70}^{-1}$ Mpc (which correspond to one-eigth and one-quarter of the simulation box size, respectively).

\begin{figure}
   \includegraphics[width=8.5cm]{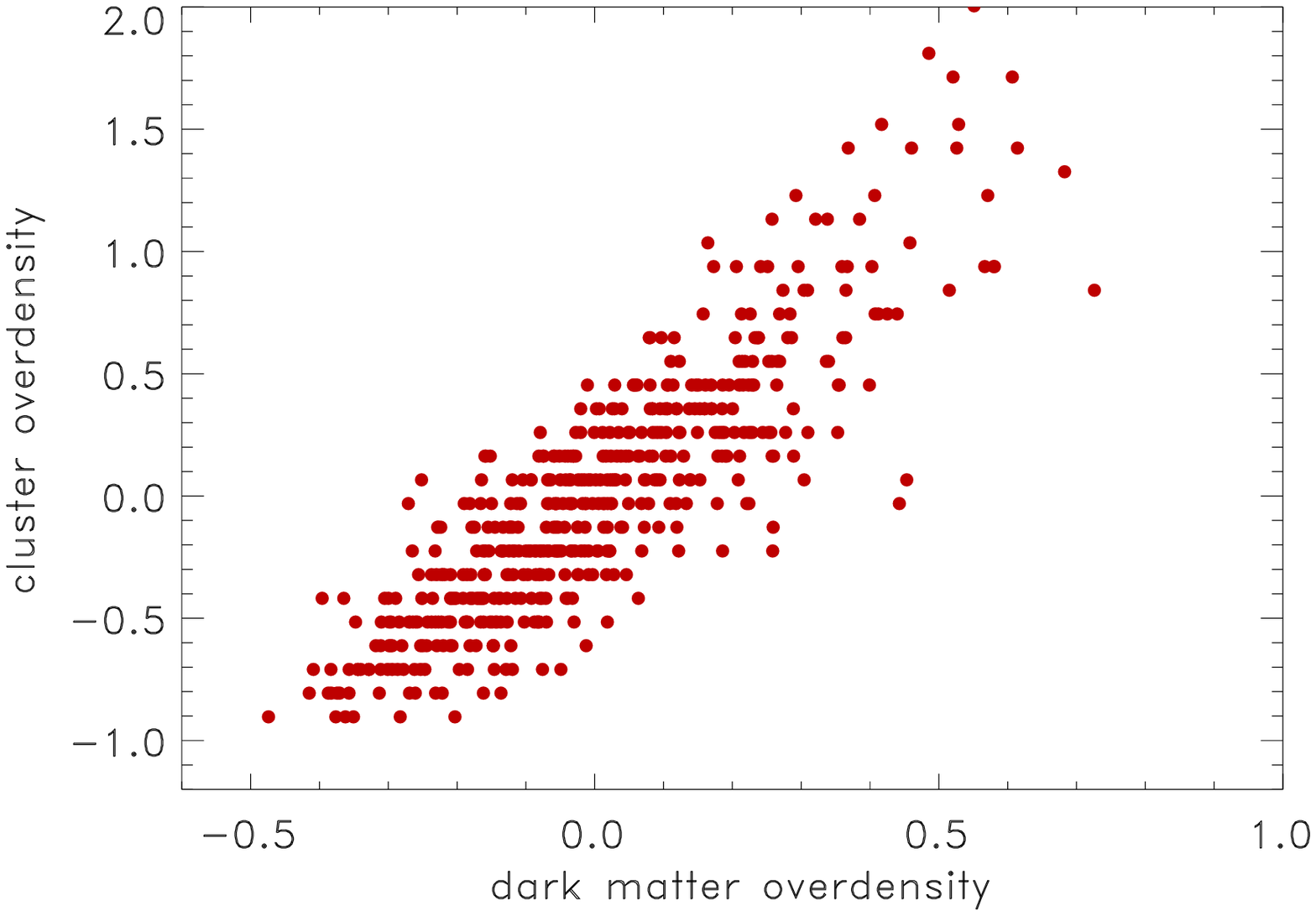}
   \includegraphics[width=8.5cm]{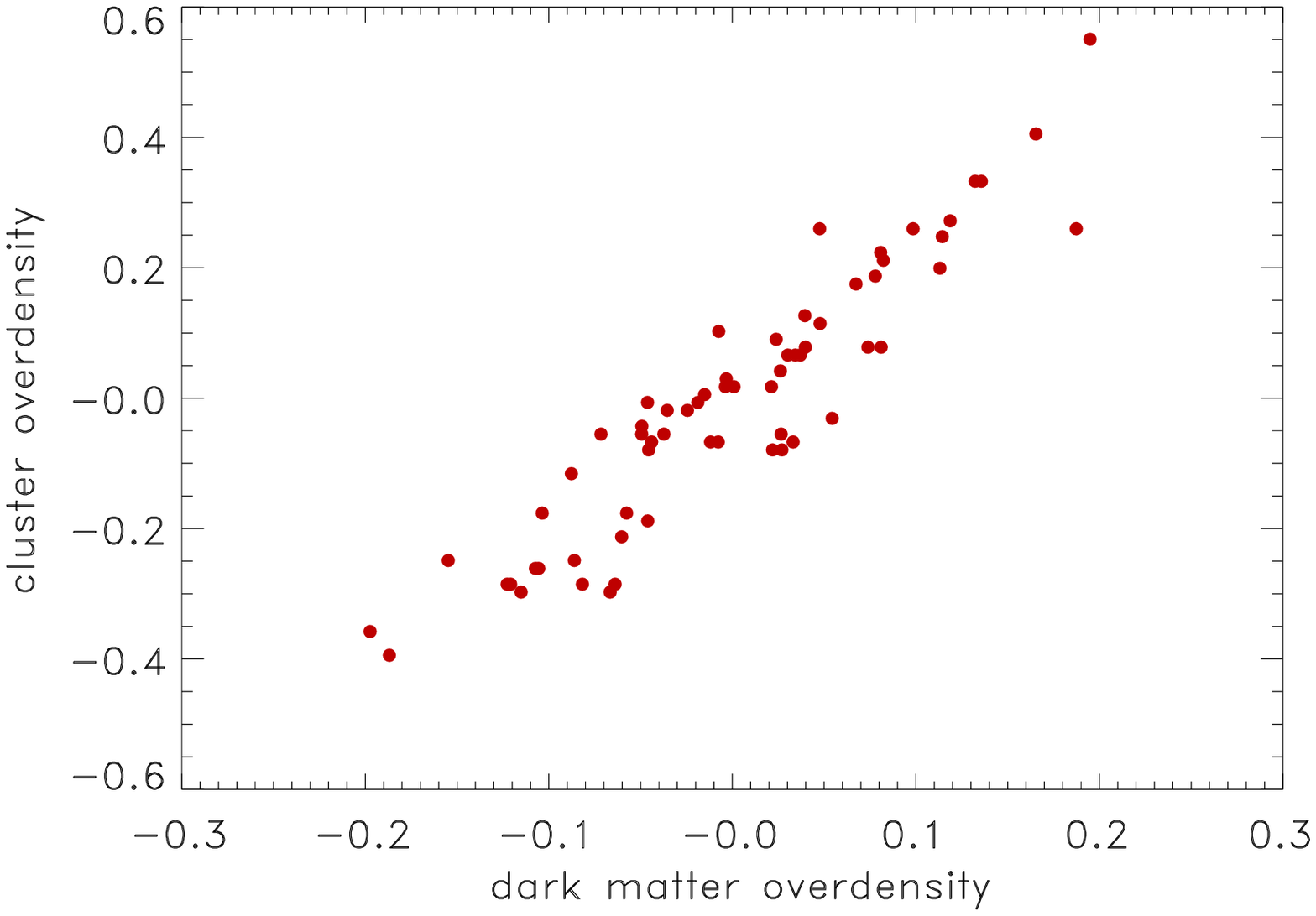}
\caption{Cluster over-/under-density with respect to the mean as
a function of the matter over-/under-density for counts in cells
with a box size of $89.3~ h_{70}^{-1}$ Mpc (upper panel) and
a size of $178.6~ h_{70}^{-1}$ Mpc (lower panel)
in the Millennium simulations.
}\label{fig_B1}
\end{figure}

The results of the two studies are presented in Fig.~\ref{fig_B1}.
What is shown is the density contrast for clusters
as a function
of the density contrast in the matter distribution. Therefore,
the slope of the relation is equal to the bias. 
We note that in both cases the distribution of clusters closely
traces that of matter. The quantitative result
important for the analysis above is the scatter
in the relation which was included in the uncertainties of the inferred
matter distribution in our analysis. 
The scatter determined for the two cases 
is $\sim 26\%$ for the smaller cells and $\sim 8\%$ for
the larger cells, which is
close to the Poisson error. In our analysis we
therefore used Poisson uncertainties.

\section{Hubble parameter as function of underdensity}

\begin{figure}
 \hskip 0.65cm
   \includegraphics[width=8.0cm]{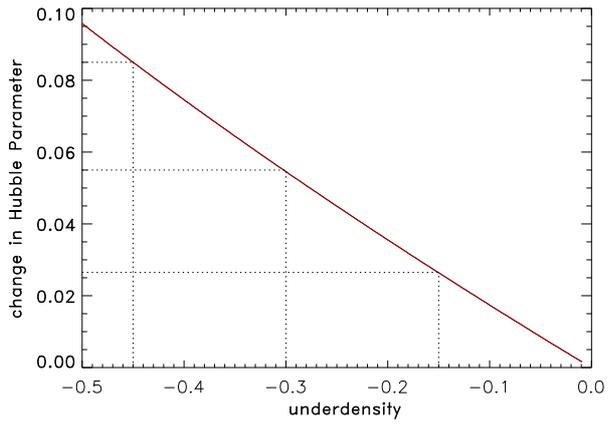}
\caption{Change of the Hubble parameter as a function of the underdensity
of the region studied. The dotted lines mark the underdensity values of
$30 \pm 15\%$. 
}\label{fig_C1}
\end{figure}

We calculated the Hubble constant that should be observed within a local
underdensity under the assumption that the underdense region is homogeneous.
Justified by Birkhoff's theorem, we integrated 
the Friedman equations from initial conditions in the early Universe 
($z = 500$) to the present time for our preferred cosmology and 
other models with sightly higher or lower densities, and compared their
expansion parameters at $z = 0$. The resulting relation between the 
underdensity and the increase of the Hubble parameter at present time 
is shown in Fig.~\ref{fig_C1}. In the literature one can      
find approximate formulas for this relation of the underdensity in
a local region and the observed Hubble constant, for example by Marra et al. (2013), which agree with our result.

\begin{thebibliography}{}

\bibitem[Agrawal (2019)]{}
Agrawal, P., Cyr-Racine, F.-Y., Pinner, D., Randall, L., 2019, arXiv:1904:01016

\bibitem[Alexander (2009)]{}
Alexander, S., Biswas, T., Notari, A., et al., 2009, JCAP, 9, 25

\bibitem[Balaguera-Antolinez (2011)]{}
Balaguera-Antolinez, A., Sanchez, A., B\"ohringer, H., et al., 2011, MNRAS, 413, 386

\bibitem[Balaguera-Antolinez (2012)]{}
Balaguera-Antolinez, A., Sanchez, A., B\"ohringer, H., et al., 2012, MNRAS, 425, 2244

\bibitem[Bardeen (1986)]{}
Bardeen, J.M., Bond, J.R., Kaiser, N., et al., 1986, ApJ, 304, 15

\bibitem[Betoule (2014)]{}
Betoule, M., Kessler, R., Guy, J., et al., 2014, A\&A, 568, A22

\bibitem[B\"ohringer (2000)]{}
B\"ohringer, H., Huchra, J.P., et al., 2000, ApJS, 129, 435

  \bibitem[B\"ohringer (2002)]{}
B\"ohringer, H., Collins, C.A., Guzzo, L., et al., 2002, ApJ, 566, 93

\bibitem[B\"ohringer (2004)]{}
B\"ohringer, H., Schuecker, P., Guzzo, L., et al., 2004, A\&A, 425, 367

\bibitem[B\"ohringer (2007)]{}
B\"ohringer, H., Schuecker, P., Pratt, G.W., et al., 2007, A\&A, 469, 363 

\bibitem[B\"ohringer (2013)]{}
B\"ohringer, H., Chon, G., Collins, C.A., et al., 2013, A\&A, 555, A30

\bibitem[B\"ohringer (2014)]{}
B\"ohringer, H., Chon, G., Collins, C.A., et al., 2014, A\&A, 570, A31

\bibitem[B\"ohringer (2015)]{}
B\"ohringer, H., Chon, G., Bristow, M., et al., 2015,  A\&A, 574, A26

\bibitem[B\"ohringer (2016)]{}
B\"ohringer, H., Chon, G., Kronberg, P.P., A\&A, 596, A22

\bibitem[B\"ohringer (2017a)]{}
B\"ohringer, H., Chon, G., Retzlaff, J., et al., 2017a, AJ, 153, 220 

\bibitem[B\"ohringer (2017b)]{}
B\"ohringer, H., Chon, G., Fukugita, M., 2017b,  A\&A, 608, A65 

\bibitem[Busswell (2004)]{}
Busswell, G.S., Shanks, T., Outram, P.J., et al., 2004, MNRAS, 354, 991

\bibitem[Celerier (2000)]{}
C\'el\'erier, M.-N., 2000, A\&A, 353, 63

\bibitem[Chon (2012)]{}
Chon, G., \& B\"ohringer, H., 2012, A\&A, 538, 35

\bibitem[Chon (2013)]{}
Chon, G., \& B\"ohringer, H.,  2013, MNRAS, 429, 3272

\bibitem[Chon (2014)]{}
Chon, G., B\"ohringer, H., Collins, C.A., et al., 2014   A\&A, 567, A144

\bibitem[Chon (2015)]{}
Chon, G., \& B\"ohringer, H. \& Zaroubi, S., 2015, A\&A, 575, L14

\bibitem[Collins (2000)]{}
Collins, C.A., Guzzo, L., B\"ohringer, H., et al., 2000, MNRAS, 319, 939

\bibitem[Conley (2007)]{}
Conley, A., Carlberg, R.G., Guy, J., 2007, ApJ, 664, L13

\bibitem[D'Eramo (2018)]{}
D'Eramo, F., Ferreira, R.Z., Notari, A., Bernal, J.L., 2018, JCAP, 11, 14

\bibitem[Desmond (2019)]{}
Desmond, H., Bhuvnesh, J., Sakstein, J., 2019, Phys. Rev. D, 100, 043537

\bibitem[DeVaucouleurs (1959)]{}
de Vaucouleurs, G., 1959, Sov. Ast., 3, 897

\bibitem[Di Valentino (2018)]{}
Di Valentino, E., Linder, E.V., Melchiori, A., 2018, Phys. Rev. D, 97, 043528

\bibitem[Dickey (1990)]{}
Dickey, J.M. \& Lockman, F.J., 1990, ARA\&A, 28, 215


\bibitem[Fleury (2017)]{}
Fleury, P., Clarkson, C., Maartens, R., 2017, JACP, 3, 62

\bibitem[Frith (2003)]{}
Frith, W.J., Busswell, G.S., Fong, R., et al. 2003, MNRAS, 345, 1049

\bibitem[Frith (2006)]{}
Frith, W.J., Metcalf, N., Shanks, T., 2006, MNRAS, 371, 1601

\bibitem[February (2010)]{}
February, S., Larena, J., Smith, M., et al., 2010, MNRAS, 405, 2231

\bibitem[Giovanelli (1999)]{}
Giovanelli, R., Dale, D.A., Haynes, M.P., 1999, ApJ, 525, 25

\bibitem[Guzzo (2009)]{}
Guzzo, L., Schuecker, P., B\"ohringer, H., et al., 2009, A\&A, 499, 357

 
\bibitem[Huang (1997)]{}
Huang, J.-S., Cowie, L.L., Gardner, J.P., et al., 1997, ApJ, 476, 12

\bibitem[Hudson (2004)]{}
Hudson, M.J., Smith, R.J., Lucey, J.R., et al., 2004, MNRAS, 352, 61

\bibitem[Huchra (2012)]{}
Huchra, J.P., Marci, L.M., Masters, K.L., et al., 2012, ApJS, 199, 26

\bibitem[Jasche (2019)]{}
Jasche, J \& Lavaux, G., 2019, A\&A, 625, A64

\bibitem[Jha (2007)]{}
Jha, S., Riess, A.G., Kirschner, R., 2007, ApJ, 659, 122

\bibitem[Kaiser (1986)]{}
Kaiser, N., 1986, MNRAS, 222, 323

\bibitem[Keenan (2013)]{}
Keenan, R.C., Barger, A.J., Cowie, L.L., 2013, ApJ, 775, 62

\bibitem[Kenworthy (2019)]{}
Kenworthy, C.D., Scolnic, D., Riess, A., 2019, ApJ,. 875, 145

\bibitem[Kerscher (2001)]{}
Kerscher, M., Mecke, K., Schuecker, P., et al., 2001, A\&A, 377, 1 

\bibitem[Lavaux (2011)]{}
Lavaux, G. \& Hudson, M.J., 2011, MNRAS, 416, 2840

\bibitem[Lewis (2000)]{}
Lewis, A., Challinor, A., Lasenby, A., 2000, ApJ, 538, L473

\bibitem[Lukovic (2019)]{}
Lukovic, V.V., Balakrishna, S.H., Vittorio, N., 2019, arXiv1907.11219

\bibitem[Macaulay (2019)]{}
Macaulay, E., Nichol, R.C., Bacon, D., et al., 2019 MNRAS, 486, 2184

\bibitem[Marra (2013)]{}
Marra, V., Amendola, L., Sawicki, I., et al., 2013, PRL, 110, 241305

\bibitem[Mo (1996)]{}
Mo, H.J. \& White, S.D.M., 1996, MNRAS, 282, 347 


\bibitem[Moss (2011)]{}
Moss, A., Zibin, J.P., Scott, D., 2011, Phys Rev D, 83, 103515

\bibitem[Odderskov (2017)]{}
Odderskov, I., Hannestad, S., Brandbyge, J., 2017, JCAP, 3, 22

\bibitem[Pandey  (2019)]{}
Pandey, K.L., Karwal, T., Das, S., 2019, arXiv:1902.10636

\bibitem[Perlmutter (1999)]{}
Perlmutter, S., Aldering, G., Goldhaber, G., et al., 1999, ApJ, 517, 565

\bibitem[Planck (2016)]{}
Planck Collaboration 2015 results XIII, 2016, A\&A, 594, A13

\bibitem[Planck (2018)]{}
Planck Collaboration 2018 results VI, arXiv:1807.06209

\bibitem[Poulin (2018)]{}
Poulin, V., Smith, T.L., Karval, T., Kamionkowski, M., 2019, PRL, 122, 1301

\bibitem[Pratt (2009)]{}
Pratt, G.W., Croston, J.H., Arnaud, M., B\"ohringer, H., 2009,
A\&A, 498, 361

\bibitem[Riess (2011)]{}
Riess, A.G., Macri, L., Casertano, S., et al., 2011, ApJ, 732, 129

\bibitem[Riess (2018a)]{}
Riess, A.G., Casertano, S., Yuan, W., et al., 2018a, AJ, 861, 126 

\bibitem[Riess (2018b)]{}
Riess, A.G., Casertano, S., Kenworthy, D'A., Scolnic, D., Marci, L., 2018b, arXiv:1810.03526

\bibitem[Riess (2019)]{}
Riess, A.G., Casertano, S., Yuan, W., Marci, L.M., Scolnic, D., 2019, arXiv:1903.07603 

\bibitem[Schmidt (1998)]{}
Schmidt, B., Suntzeff, N.B., Phillips, M.M., et al., 1998, ApJ, 507, 46 

\bibitem[Schuecker (2001)]{}
Schuecker, P., B\"ohringer, H., Guzzo, L., et al., 
2001, A\&A,368, 86

\bibitem[Schuecker (2002)]{}
Schuecker, P., Guzzo, L., Collins, C.A., et al., 2002,
MNRAS, 335, 807

\bibitem[Schuecker (2003a)]{}
Schuecker, P., B\"ohringer, H., Collins, C.A. et al., 2003a, 
A\&A, 398, 867

\bibitem[Schuecker (2003b)]{}
Schuecker, P., Caldwell, R.R., B\"ohringer, H., et al., 2003b, A\&A, 402, 53

\bibitem[Scolnic (2018)]{}
Scolnic, D.M., Jones, D.O., Rest, A., et al., 2018, ApJ, 859, 101

\bibitem[Shanks (2019a)]{}
Shanks, T., Hogarth, L.M., Metclafe, N., 2019a, MNRAS, 484, L64

\bibitem[Shanks (2019b)]{}
Shanks, T., Hogarth, L.M., Metclafe, N., et al., 2019b, arXiv:1909.01878

\bibitem[Sheth (1999)]{}
Sheth, R.K. \& Tormen, G., 1999, MNRAS, 308,  119 

\bibitem[Tinker (2010)]{}
Tinker, J.L., Robertson, B.E., Kravtsov. A.V., 2010, ApJ, 724, 878
 
\bibitem[Tomita (2000)]{}
Tomita, K., 2000, MNRAS, 326, 287

\bibitem[Tomita (2001)]{}
Tomita, K., 2001, Prog. Theor. Phys. 106, 929

\bibitem[Tr\"umper (1993)]{}
Tr\"umper, J., 1993, Science, 260, 1769

\bibitem[Tully (2016)]{}
Tully, R.B., Courtois, H.M., Sorce, J.G., 2016, AJ, 152, 50


\bibitem[Tully (2019)]{}
Tully, R.B., Pomarede, D., Graziani, R., et al., 2019, ApJ, 880, 24


\bibitem[Vattis (2019)]{}
Vattis, K., Koushiappas, S.M., Loeb, A., 2019, arXiv:1903.06220

\bibitem[Voges (1999)]{}
Voges, W., Aschenbach, B., Boller, T., et al. 1999,  A\&A, 349, 389

\bibitem[Whitbourn (2014)]{}
Whitbourn, J.R. \& Shanks, T., 2014, MNRAS, 437, 2146

\bibitem[Wojtak (2014)]{}
Wojtak, R., Knebe, A., Watson, W.A., 2014, MNRAS, 438, 1805

\bibitem[Wu (2017)]{}
Wu, H.-Y.  \& Huterer, D., 2017, MNRAS, 471, 4946

\bibitem[Yu (2013)]{}
Yu, B., 2013, JCAP, 3, 13

\bibitem[Zehavi (1998)]{}
Zehavi, I., Riess, A.G., Kirschner, R., et al., 1998, ApJ, 503, 483

\end{thebibliography}
\end{document}